\input harvmac
\noblackbox
\newcount\figno
\figno=0
\def\fig#1#2#3{
\par\begingroup\parindent=0pt\leftskip=1cm\rightskip=1cm\parindent=0pt
\baselineskip=11pt
\global\advance\figno by 1
\midinsert
\epsfxsize=#3
\centerline{\epsfbox{#2}}
\vskip 12pt
\centerline{{\bf Figure \the\figno:} #1}\par
\endinsert\endgroup\par}
\def\figlabel#1{\xdef#1{\the\figno}}

\def\np#1#2#3{Nucl. Phys. {\bf B#1} (#2) #3}
\def\pl#1#2#3{Phys. Lett. {\bf B#1} (#2) #3}
\def\prl#1#2#3{Phys. Rev. Lett.{\bf #1} (#2) #3}
\def\physrev#1#2#3{Phys. Rev. {\bf D#1} (#2) #3}


\font\cmss=cmss10
\font\cmsss=cmss10 at 7pt
\def\rlx{\relax\leavevmode}
\def\inbar{\vrule height1.5ex width.4pt depth0pt}
\def\IC{\relax\,\hbox{$\inbar\kern-.3em{\rm C}$}}
\def\IN{\relax{\rm I\kern-.18em N}}
\def\IP{\relax{\rm I\kern-.18em P}}
\def\ZZ{\rlx\leavevmode\ifmmode\mathchoice{\hbox{\cmss Z\kern-.4em Z}}
 {\hbox{\cmss Z\kern-.4em Z}}{\lower.9pt\hbox{\cmsss Z\kern-.36em Z}}
 {\lower1.2pt\hbox{\cmsss Z\kern-.36em Z}}\else{\cmss Z\kern-.4em
 Z}\fi}
\def\IZ{\relax\ifmmode\mathchoice
{\hbox{\cmss Z\kern-.4em Z}}{\hbox{\cmss Z\kern-.4em Z}}
{\lower.9pt\hbox{\cmsss Z\kern-.4em Z}}
{\lower1.2pt\hbox{\cmsss Z\kern-.4em Z}}\else{\cmss Z\kern-.4em
Z}\fi}

\def\narrowplus{\kern -.04truein + \kern -.03truein}
\def\narrowminus{- \kern -.04truein}
\def\narrowminussub{\kern -.02truein - \kern -.01truein}

\def\kh{K\"{a}hler }

\def\g{{\gamma}}

\def\r{{\rightarrow}}

\def\frac#1#2{{#1\over #2}}

\def\CM{{\cal M}}

\def\IZ{\relax\ifmmode\mathchoice
{\hbox{\cmss Z\kern-.4em Z}}{\hbox{\cmss Z\kern-.4em Z}}
{\lower.9pt\hbox{\cmsss Z\kern-.4em Z}}
{\lower1.2pt\hbox{\cmsss Z\kern-.4em Z}}\else{\cmss Z\kern-.4em
Z}\fi}
\def\IB{\relax{\rm I\kern-.18em B}}
\def\IC{{\relax\hbox{$\inbar\kern-.3em{\rm C}$}}}
\def\ID{\relax{\rm I\kern-.18em D}}
\def\IE{\relax{\rm I\kern-.18em E}}
\def\IF{\relax{\rm I\kern-.18em F}}
\def\IG{\relax\hbox{$\inbar\kern-.3em{\rm G}$}}
\def\IGa{\relax\hbox{${\rm I}\kern-.18em\Gamma$}}
\def\IH{\relax{\rm I\kern-.18em H}}
\def\II{\relax{\rm I\kern-.18em I}}
\def\IK{\relax{\rm I\kern-.18em K}}
\def\IP{\relax{\rm I\kern-.18em P}}

\def\p{\phi}

\font\cmss=cmss10 \font\cmsss=cmss10 at 7pt
\def\IR{\relax{\rm I\kern-.18em R}}

\def\S{{\Sigma}}
\def\tp{{\tilde \phi}}

%

%
%
\def\eqnn#1{\xdef #1{(\secsym\the\meqno)}\writedef{#1\leftbracket#1}%
\global\advance\meqno by1\wrlabeL#1}
\def\eqna#1{\xdef #1##1{\hbox{$(\secsym\the\meqno##1)$}}
\writedef{#1\numbersign1\leftbracket#1{\numbersign1}}%
\global\advance\meqno by1\wrlabeL{#1$\{\}$}}
\def\eqn#1#2{\xdef #1{(\secsym\the\meqno)}\writedef{#1\leftbracket#1}%
\global\advance\meqno by1$$#2\eqno#1\eqlabeL#1$$}

\lref\rpol{J. Polchinski, ``TASI Lectures on D-Branes,''
hep-th/9611050\semi J. Polchinski, S. Chaudhuri and C. Johnson,
``Notes on D-Branes,'' hep-th/9602052. }
\lref\rBFSS{T. Banks, W. Fischler, S. H. Shenker, and L. Susskind, ``M
Theory As A Matrix Model: A Conjecture,''
hep-th/9610043, Phys. Rev. {\bf D55} (1997) 5112.}
\lref\rwtensor{E. Witten, ``Some Comments on String Dynamics,''
hep-th/9507121.}
\lref\rstensor{A. Strominger, ``Open P-Branes,'' hep-th/9512059,
\pl{383}{1996}{44}.}
\lref\rsdecoupled{N. Seiberg, ``New Theories in Six Dimensions and
Matrix Description of M-theory on $T^5$ and $T^5/\IZ_2$,''
hep-th/9705221.}
\lref\rashoke{ A. Sen, ``Kaluza-Klein Dyons in String
Theory,'' hep-th/9705212; ``A Note on Enhanced Gauge Symmetries in
M and String Theory,'' hep-th/9707123; ``Dynamics of Multiple
Kaluza-Klein Monopoles in M and String Theory,'' hep-th/9707042.}
\lref\rtwoform{J. P. Gauntlett and D. Lowe, ``Dyons and S-Duality in
N=4 Supersymmetric Gauge Theory,'' hep-th/9601085,
\np{472}{1996}{194}\semi K. Lee, E. Weinberg and P. Yi,
``Electromagnetic Duality and $SU(3)$ Monopoles,'' hep-th/9601097,
\pl{376}{1996}{97}.}
\lref\rmoore{D. Berenstein, R. Corrado and J. Distler, ``On the Moduli Spaces
of M(atrix)-Theory Compactifications,'' hep-th/9704087\semi  S. Elitzur, A.
Giveon, D. Kutasov and E. Rabinovici, ``Algebraic Aspects of Matrix Theory on
$T^d$,'' hep-th/9707217\semi A. Losev, G. Moore, and S. Shatashvili, ``M \& m's
,''
hep-th/9707250\semi I. Brunner and A. Karch, ``Matrix Description of
M-theory on $T^6$,'' hep-th/9707259.}
\lref\rHG{ A. Hanany and G. Lifschytz, ``M(atrix) Theory on $T^6$
and a m(atrix) Theory Description of KK Monopoles,'' hep-th/9708037.}
\lref\rDVV{R. Dijkgraaf, E. Verlinde and H. Verlinde, ``BPS Spectrum
of the Five-Brane and Black Hole Entropy,'' hep-th/9603126,
\np{486}{1997}{77}; ``BPS Quantization of the Five-Brane,''
hep-th/9604055, \np{486}{1997}{89}; ``5D Black Holes and Matrix Strings,''
 hep-th/9704018.}
\lref\rsixbrane{P. Townsend, ``The Eleven Dimensional Supermembrane
Revisited,'' hep-th/9501068, \pl{350}{1995}{184}.}
\lref\rmultitn{S. Hawking, ``Gravitational Instantons,''
Phys. Lett. {\bf 60A} (1977) 81\semi
G. Gibbons and S. Hawking, ``Classification of Gravitational Instanton
Symmetries,'' Comm. Math. Phys. {\bf 66} (1979) 291\semi
R. Sorkin, ``Kaluza-Klein Monopole,''
\prl{51}{1983}{87}\semi D. Gross and M. Perry, ``Magnetic Monopoles in
Kaluza-Klein Theories,'' \np{226}{1983}{29}.}
\lref\rmIIB{P. Aspinwall, ``Some Relationships Between Dualities in
String Theory,'' hep-th/9508154, Nucl. Phys. Proc. Suppl. {\bf 46}
(1996) 30\semi J. Schwarz, ``The Power of M Theory,''
hep-th/9510086, \pl{367}{1996}{97}. }
\lref\rgeneralsixbrane{ C. Hull,  ``Gravitational Duality, Branes and
Charges,'' hep-th/9705162\semi E. Bergshoeff, B. Janssen, and
T. Ortin, ``Kaluza-Klein Monopoles and Gauged Sigma Models,''
hep-th/9706117\semi Y. Imamura, ``Born-Infeld Action and Chern-Simons
Term {}from Kaluza-Klein Monopole in M-theory,'' hep-th/9706144\semi
R. Gregory, J. Harvey and G. Moore, ``Unwinding strings and T-duality of
 Kaluza-Klein and H-Monopoles,'' hep-th/9708086.}
\lref\rbd{M. Berkooz and M. Douglas, ``Five-branes in M(atrix)
Theory,'' hep-th/9610236, \pl{395}{1997}{196}.}
\lref\rbraneswith{M. Douglas, ``Branes within Branes,''
hep-th/9512077.}
\lref\rfourfive{E. Witten, ``Solutions Of Four-Dimensional Field Theories Via M
Theory, '' hep-th/9703166.}
\lref\rquantumfive{O. Aharony, M. Berkooz, S. Kachru, N. Seiberg, and
E. Silverstein, ``Matrix Description of Interacting Theories in Six
Dimensions,'' hep-th/9707079.}
\lref\rstringfive{E. Witten, ``On The Conformal Field Theory of The
Higgs Branch,'' hep-th/9707093.}
\lref\rSS{S. Sethi and L. Susskind, ``Rotational Invariance in the
M(atrix) Formulation of Type IIB Theory,'' hep-th/9702101,
\pl{400}{1997}{265}.}
\lref\rBS{T. Banks and N. Seiberg, ``Strings from Matrices,''
hep-th/9702187, \np{497}{1997}{41}.}
\lref\rreview{N. Seiberg, ``Notes on Theories with 16 Supercharges,''
hep-th/9705117.}
\lref\rDVVstring{R. Dijkgraaf, E. Verlinde and H. Verlinde, ``Matrix
String Theory,'' hep-th/9703030.}
\lref\rprobes{M. Douglas, ``Gauge Fields and D-branes,''
hep-th/9604198.}
\lref\rdoumoo{M. Douglas and G. Moore, ``D-Branes,
Quivers, and ALE Instantons,'' hep-th/9603167.}
\lref\rfischler{M. Douglas, ``Enhanced Gauge Symmetry in M(atrix)
Theory,'' hep-th/9612126\semi
W. Fischler and A. Rajaraman, ``M(atrix) String Theory
on K3,'' hep-th/9704123.}
\lref\rDEcases{C. Johnson and R. Myers, ``Aspects of Type IIB Theory on ALE
Spaces,'' hep-th/9610140, \physrev{55}{1997}{6382}.}
\lref\rDG{D.-E. Diaconescu and J. Gomis, ``Duality in Matrix Theory
and Three Dimensional Mirror Symmetry,'' hep-th/9707019.}
\lref\rsprobes{N. Seiberg, ``Gauge Dynamics And Compactification To
Three Dimensions,'' hep-th/9607163, \pl{384}{1996}{81}}
\lref\rswthree{N. Seiberg and E. Witten, ``Gauge Dynamics and
Compactifications to Three Dimensions,'' hep-th/9607163.}
\lref\rthroat{D.-E. Diaconescu and N. Seiberg, ``The Coulomb Branch of
$(4,4)$ Supersymmetric Field Theories in Two Dimensions,''
hep-th/9707158. }
\lref\rTduality{T. Banks, M. Dine, H. Dykstra and W. Fischler,
``Magnetic Monopole Solutions of String Theory,''
\pl{212}{1988}{45}\semi C. Hull and P. Townsend, ``Unity of
Superstring Dualities,'' hep-th/9410167, \np{438}{109}{1995}. }
\lref\rvafa{H. Ooguri, C. Vafa, ``Two Dimensional Black Hole and Singularities
of
Calabi-Yau Manifolds,'' Nucl.Phys. {\bf B463} (1996) 55,
hep-th/9511164\semi D. Kutasov, ``Orbifolds and Solitons,'' Phys. Lett
{\bf B383} (1996) 48, hep-th/9512145.}
\lref\raps{P. Argyres, R. Plesser and N. Seiberg, ``The Moduli Space
of N=2 SUSY QCD and Duality in N=1 SUSY QCD,'' hep-th/9603042,
\np{471}{1996}{159}.}
\lref\rgms{O. Ganor, D. Morrison and N. Seiberg, ``Branes, Calabi-Yau
Spaces, and Toroidal Compactification of the N=1 Six Dimensional $E_8$
Theory,'' hep-th/9610251, \np{487}{1997}{93}.}
\lref\rchs{ C.G. Callan, J.A. Harvey, A. Strominger, ``Supersymmetric
String Solitons,'' hep-th/9112030, \np{359}{1991}{611}\semi S.-J. Rey,
in ``The  Proc. of the Tuscaloosa Workshop
1989,'' 291; Phys. Rev. {\bf D43} (1991) 526; S.-J. Rey, In DPF Conf.
1991, 876.}
\lref\rmotl{L. Motl, ``Proposals on nonperturbative superstring
interactions,'' hep-th/9701025.}
\lref\rwati{W. Taylor IV, ``D-Brane Field Theory on Compact Space,''
hep-th/9611042, \pl{394}{1997}{283}.}
\lref\rkin{K. Intriligator, ``New String Theories in Six Dimensions via Branes
at Orbifold Singularities,'' hep-th/9708117.}
\lref\rns{N. Seiberg and S. Sethi, ``Comments on Neveu-Schwarz Five-Branes, ''
hep-th/9708085.}
\lref\rlowe{D. Lowe, ``$E_8 \times E_8$ Small Instantons in Matrix Theory,''
hep-th/9709015\semi O. Aharony, M. Berkooz, S. Kachru and E. Silverstein,
``Matrix Description of (1,0) Theories in Six Dimensions,'' hep-th/9709118. }
\lref\rmirror{K. Intriligator and N. Seiberg, ``Mirror Symmetry in Three
Dimensional Gauge Theories,'' hep-th/9607207, \pl{386}{1996}{513}. }
\lref\rtoappear{E. Witten, to appear. }
\lref\rdlcq{L. Susskind, ``Another Conjecture about M(atrix) Theory,''
hep-th/9704080.}
\lref\rconjecture{J. de Boer, K. Hori, H. Ooguri and Y. Oz, ``Mirror Symmetry
in Three-Dimensional Gauge Theories, Quivers and D-branes,'' hep-th/9611063,
\np{493}{1997}{101}.}
\lref\rhanany{A. Hanany and E. Witten, ``Type IIB Superstrings, BPS Monopoles,
And Three-Dimensional Gauge Dynamics,'' hep-th/9611230, \np{492}{1997}{152}. }
\lref\ros{O. Ganor and S. Sethi, to appear. }
\lref\rori{O. Ganor, S. Ramgoolam and W. Taylor IV, ``Branes, Fluxes and
Duality in M(atrix)-Theory,'' hep-th/9611202, \np{492}{1997}{191}.}
\lref\rhmon{S. Sethi and M. Stern, ``A Comment on the Spectrum of
H-Monopoles,'' hep-th/9607145, \pl{398}{1997}{47}.}
\lref\rglobal{D.-E. Diaconescu and R. Entin, ``A Non-Renormalization Theorem
for the d=1, N=8 Vector Multiplet,'' hep-th/9706059.}
\lref\rDE{C. Johnson and R. Myers, ``Aspects of Type IIB Theory on ALE
Spaces,''hep-th/9610140, \physrev{55}{1997}{6382}.}
\lref\rbrody{J. Brodie, ``Two Dimensional Mirror Symmetry from M-theory,''
hep-th/9709228.}
\lref\rconvmirror{B. Greene and R. Plesser, ``Duality in Calabi-Yau Moduli
Space, '' \np{338}{1990}{15}\semi W. Lerche, C. Vafa and N. Warner, ``Chiral
Rings in N=2 Superconformal Theories,'' \np{324}{1989}{427}.  }
\lref\rks{A. Kapustin and S. Sethi, to appear. }

\Title{\vbox{\hbox{hep-th/9710005}\hbox{IASSNS-HEP-97/103}}}
{The Matrix Formulation of Type IIB Five-Branes}

\smallskip
\centerline{Savdeep Sethi\footnote{$^\ast$}{sethi@sns.ias.edu} }
\medskip\centerline{\it School of Natural Sciences}
\centerline{\it Institute for Advanced Study}\centerline{\it
Princeton, NJ 08540, USA}


\vskip 1in

We present a matrix model which interpolates between type IIA and type IIB NS
five-branes. The matrix description involves a three-dimensional bulk quantum
field theory interacting with impurities localized in one spatial direction. We
obtain a dual matrix formulation for the exotic six-dimensional theory on
coincident type IIB NS five-branes by studying the T-dual description in terms
of Kaluza-Klein monopoles in type IIA string theory. After decoupling the bulk
physics, the matrix description reduces to the conformal field theory of the
Coulomb branch for the type IIA matrix string propagating on certain singular
spaces. In many ways, this dual realization of superconformal theories is the
two-dimensional analogue of three-dimensional mirror symmetry.

\vskip 0.1in
\Date{9/97}

\newsec{Introduction}

Exotic six-dimensional theories have received increased attention recently. The
most studied of these theories is the $(2,0)$ field theory, which was first
discovered in \refs{\rwtensor, \rstensor}. This field theory is an interacting
superconformal theory living on $k$ coincident five-branes in M theory. In
large part, the reason this theory is so interesting is that it remains
interacting in the limit where the eleven-dimensional Planck constant,
$M_{pl}$, is taken to infinity. In this limit, the theory living on the
five-branes decouples from the bulk spacetime modes, leaving a consistent,
complete theory in six-dimensions. This type of decoupling argument has been
applied to the five-branes in string theory to argue for the existence of new
six-dimensional theories \rsdecoupled. The theory on $k$ parallel branes is
associated to an $A_{k-1}$ singularity, but similar six-dimensional theories
can be associated to $D$ and $E$ singularities. Aspects of the $(2,0)$ field
theory associated to $D$ and $E$ singularities have been studied in
\refs{\rwtensor, \rreview}.

The aim of this paper is to consider the case of six-dimensional theories with
$(1,1)$ supersymmetry. As before, the theories associated to $A_{k-1}$
singularities can be realized in terms of coincident NS five-branes in type IIB
string theory. The existence of complete theories on type IIB NS five-branes
has been argued in \rsdecoupled. The fields in a theory with $(1,1)$
supersymmetry in six-dimensions and no gravity must appear in vector
supermultiplets. At low-energies, the theory on type IIB five-branes is
therefore a gauge theory. The gauge coupling for the $U(k)$ gauge theory living
on $k$ parallel branes depends only on the string scale, $M_s$ \rsdecoupled.
Since the gauge coupling is independent of the type IIB string coupling,
$g_s^B$, the theory on the branes remains interacting in the limit where the
string coupling vanishes and $M_{pl}\r \infty$. This is to be constrasted with
the theories of tensor multiplets, where the self-duality constraint on the
three-form field strength guarantees that these theories remain interacting
when $M_{pl} \r \infty$. In addition to the vector particles, the
six-dimensional theory includes strings of tension $M_s^2$, which can be viewed
as bound states of fundamental strings with the NS five-branes, or alternately,
as instantons of the $U(k)$ gauge theory. Our goal is to find a matrix
description of five-branes in type IIB string theory. From this matrix
description, we can extract a matrix definition for the exotic $(1,1)$ theory
associated to coincident NS five-branes. We will extend the definition to the
theories associated to general $A-D-E$ singularities. The decoupling limit for
matrix theories involving five-branes in M theory, type IIA and heterotic
string theory has been discussed recently in
\refs{\rquantumfive\rstringfive\rns\rkin -\rlowe}, partly motivated by earlier
work \rDVV.

In obtaining a matrix description for these interacting six-dimensional
theories, we will find dual realizations of certain superconformal field
theories with eight supersymmetries in two dimensions. One realization is a
limit of a $2+1$-dimensional theory with impurities localized in one space
dimension. The bulk theory has sixteen supersymmetries, but the impurities
break half the supersymmetries. The second realization is in terms of a
$1+1$-dimensional theory of vector and hypermultiplets with $(4,4)$
supersymmetry. In many ways, this dual description of certain superconformal
fixed points is the two-dimensional analogue of the three-dimensional mirror
symmetry found by Intriligator and Seiberg \rmirror. Observations related to
those presented here have been made by Witten \rtoappear. As I completed this
project, related work appeared in \rbrody.

\newsec{Matrix Models for Type IIB Five-Branes}

\subsec{The M theory construction of type IIB five-branes}

Let us begin by considering the M theory construction of five-branes in type
IIB string theory.  M theory compactified on a two-torus is equivalent to the
type IIB string on a circle whose size grows inversely with the volume of the
two-torus \rmIIB. The torus, $T^2$, on which we are compactifying is defined by
its complex structure, $\tau$, and its volume. For simplicity, let us take the
torus to be rectangular with sides of length, $2\pi R_1$ and $2\pi R_2$. Since
we will eventually restrict to NS five-branes, this choice of complex structure
will not affect the following discussion in a significant way. The radius of
the type IIB circle is,
\eqn\IIBcircle{ R_B = {1\over M_{pl}^3 R_1 R_2} = {1\over M_s^2 R_2}. }
The string scale is defined in terms of $M_{pl}$ and $R_1$ by the relation:
\eqn\msfi{M_s^2= R_1 M_{pl}^3. }
The type IIB string coupling is given by, $g_s^B = R_1/R_2$. There are now two
ways to realize a $(p,q)$ five-brane in the type IIB theory. The first
realization is in terms of an M theory five-brane wrapped on a $(p,q)$ cycle of
the two-torus. This corresponds to a $(p,q)$ five-brane in type IIB string
theory wrapped on the circle with radius $R_B$. Note that in these conventions,
an M theory five-brane wrapped on $R_2$ results in a type IIB NS five-brane.
When the area of the torus is taken to zero with $M_s$ held fixed, we obtain a
$(p,q)$ five-brane in flat non-compact space.

The second realization is obtained by wrapping an M theory six-brane along a
$(p,q)$ cycle of the torus. This corresponds to a $(p,q)$ five-brane in the
type IIB theory with a compact transverse dimension of radius $R_B$. Again,
taking the area of the torus to zero results in a $(p,q)$ five-brane in flat
non-compact spacetime. We will examine the matrix realization for this second
case in a limit that captures the decoupled physics in the following section.
Since both ways of constructing a $(p,q)$ five-brane must be equivalent, the
corresponding matrix descriptions of the decoupled theory must also be
equivalent. This equivalence gives a two-dimensional analogue of the
three-dimensional mirror symmetry found by Intriligator and Seiberg \rmirror.

The limit in which the bulk physics decouples from the six-dimensional theory
on the five-branes has been described in \rsdecoupled. We need to take,
\eqn\delimits{\eqalign{
&M_{pl} \rightarrow \infty, \cr
& g_s^B \rightarrow 0, \cr}}
while holding fixed $M_s$.  We should then find a family of new theories
parametrized by $M_s$. In case one, we also need to take the area of the torus
to zero to obtain a five-brane with no compact internal dimensions. For case
two, the size of the transverse circle is irrelevant in the decoupling limit
\rns. In this case, the moduli space for the gauge theory on $k$ coincident
branes might seem to be,
\eqn\modsf{ \CM = {(\IR^3 \times S^1)^k\over {\bf S}_k},}
with compact directions. In actuality, the period for the compact scalars
contains a factor of the inverse string coupling and so decompactifies in the
decoupling limit. We are therefore free to take $R_B$ to any convenient value.
We shall exploit this freedom to solve for the matrix description of the theory
on NS five-branes.

\subsec{NS five-branes in type IIB matrix theory}

Let us turn to the matrix description of case one. This involves a rather
interesting, slightly novel theory, which is similar in many ways to systems
with impurities which are studied in  condensed matter physics. Similar systems
appear in matrix descriptions of Yang-Mills theories in less than six
space-time dimensions, but that is a topic to be explored elsewhere \ros. The
matrix theory for the type IIB string without any five-branes has been
considered in \refs{\rSS, \rBS}, generalizing the original matrix conjecture
for uncompactified M theory \rBFSS. The parameters of the matrix theory are
fixed in terms of the size of the longitudinal direction, $R$, and the two
radii of the torus, $R_1$ and $R_2$. The matrix description is a
$2+1$-dimensional $U(N)$ gauge theory with sixteen supersymmetries. This gauge
theory will be our bulk theory. The parameter $N$ is the number of zero-branes.
At least for cases with sixteen supersymmetries, the parameter $N$ need not be
taken to infinity if the matrix model is viewed as describing M theory
quantized in the discrete light-cone formalism \rdlcq. The gauge theory lives
on a torus with radii,
\eqn\ymdim{ \Sigma_i = {1 \over M_{pl}^3 R_i R},}
and the coupling constant for the bulk gauge theory is given by
\eqn\ymcoupling{ g^2_{YM} = {R \over R_1 R_2} =  R^3
M_{pl}^6 \S_1 \S_2. }
For studying the decoupling limit, it is convenient to express these parameters
in string units where,
\eqn\newymdim{ \eqalign{ \S_1 &= {1 \over M_s^2 R} \cr
                         \S_2 &= {1\over M_s^2 R} g_s^B.}}
The presense of parallel M theory five-branes breaks half of the
supersymmetries in the Yang-Mills theory. The matrix theory for the case where
the $k$ five-branes are placed transverse to the two-torus has been discussed
in \rns. It is a generalization of the quantum mechanics for a longitudinal
five-brane which describes the coupling of zero-branes to a background type IIA
four-brane \rbd. In this situation, the five-branes are represented by $k$
hypermultiplets in the fundamental of the gauge group. These $2+1$-dimensional
hypermultiplets come from the quantum mechanical hypermultiplets representing
the $0-4$ strings. The quantum mechanical degrees of freedom are promoted to
fields because of the two compact transverse dimensions \refs{\rwati, \rBFSS}.
The spacetime physics is encoded in the structure of the Coulomb branch which,
as a function of $R_2$,  interpolates between the metric for the type IIB
Kaluza-Klein monopole and the tube metric of the type IIA NS five-brane \rchs.
When there is more than one coincident brane, there is also a Higgs branch
which describes the physics localized on the five-branes. It is interesting to
note that these theories have been studied in the context of three-dimensional
mirror symmetry \rmirror, where candidate duals have been conjectured
\rconjecture.

In our case, we want type IIB NS five-branes, so we consider $k$ longitudinal M
theory five-branes wrapped on $R_2$. After T-duality on $R_2$, the $0-4$
strings become hypermultiplets located at points points on $\S_2$. The $N$
zero-branes become D-strings wrapping $\S_2$. A further T-duality on $R_1$
promotes the $0-4$ strings to $1+1$-dimensional hypermultiplets located at
points on $\S_2$. The $N$ zero-branes become two-branes wrapping the dual
torus. From the perspective of the bulk theory, the hypermultiplets are
localized impurities which break half of the supersymmetries. The position of
the $k$ points on $\S_2$ is determined by the choice of flat connection for the
$U(k)$ gauge-field living on the wrapped four-brane system. Generally, when the
four-branes are not coincident, some of the hypermultiplets are massive.

 After this sequence of T-dualities, this theory becomes one containing $N$
two-branes wrapped on $T^2$ with $k$ four-branes located at points on $\S_2$,
and wrapped along the $\S_1$ direction. When there is more than one four-brane,
the two-branes can `break' at the positions of the wrapped four-branes in a way
which is essentially T-dual to the breaking of D-strings wrapped on a circle at
the location of three-branes located at points on the circle. When the bulk
gauge coupling is taken to infinity, this type IIA configuration of four-branes
and two-branes is better described in M theory since the eleventh dimension
decompactifies. This is analogous to the situation in \rfourfive, but here the
result is a system of five-branes with membranes stretched between them.

For simplicity, let us place the $k$ longitudinal five-branes at the same point
in the transverse space,
$$x_6= \ldots = x_9=0.$$
We can place $k_p$ five-branes at the point $x_2=x_p$, where $x_2$ is a
coordinate for the circle with radius $\S_2$ and $\sum k_p=k$. The bulk
Yang-Mills theory has seven scalars, $\p_i$, in the adjoint of $U(N)$ which
transform under a global $Spin(7)$ symmetry. In the presence of the
five-branes, the global symmetry decomposes into $ Spin(3) \times Spin(4)$
where the $Spin(3)$ rotates $\p_3,\p_4,\p_5$, while $Spin(4)$ acts on $\p_6 ,
\p_7, \p_8, \p_9$.  These latter scalars parametrize motion away from the
five-branes. Since we are interested in the decoupled physics, we will freeze
the expectation values for these fields at the position of the longitudinal
five-branes. Since the spatial directions are compact, the notion of a
well-defined expectation value exists only classically. The Lagragian for this
system of impurities interacting with the bulk modes can be obtained from the
original $0-0$ and $0-4$ lowest string modes using an extension of the method
described in \refs{\rwati, \rBFSS, \rori}. We start with the positions of N
zero-branes in $\IR^9$ described by nine matrices, $ X^i$, in the adjoint of
$U(N)$. In addition, we have $k$ hypermultiplets $Q^f_A$ which are complex
bosons in the fundamental of $U(N)$. The flavor index, $f=1,\ldots,k$ while
$A=1,2$ is an index for the doublet representation of $SU(2)_R$. Compactifying
$X^1$ is accomplished by taking an array of $0-4$ systems consisting of the
original system and all its translates shifted along $X^1$ by $ 2\pi n \S_1$
for all integer $n$. The gauge group then becomes infinite-dimensional, and the
$X^i$ for $i>1$ and hypermultiplets $Q^f$ can be organized into fields
depending on the periodic coordinate $x_1$. Lastly, $X^1$ can be replaced by
the connection, $ - i\partial_1 - A_1(x_1)$. In this way, we obtain
$1+1$-dimensional Yang-Mills coupled to $k$ hypermultiplets. A similar story
occurs for the $X^i$ fields when $X^2$ is compactified. The $X^i$ for $i>2$
become $2+1$-dimensional fields $\p_i$ depending on $x_1, x_2$ while $X^2$ is
replaced by the connection, $-i\partial_2 - A_2$. These bulk fields have a
standard Lagrangian whose bosonic part is,
\eqn\bulklag{ L_{\rm bulk} = \int{dt dx_1dx_2 \, \, \left(- {1\over 4} \Tr (
F_{\mu\nu}^2 ) - {1\over 2} \Tr ( D_\mu \p_i)^2 - {1\over 2} \sum_{i<j} \Tr
[\p_i, \p_j]^2 \right).}}
However, the hypermultiplets now need to be treated differently. Let us take
the $k_p$ hypermultiplets located at $x_p$. There are similar expressions for
the remaining localized hypermultiplets.  The terms in the Lagrangian that come
from the reduction of the six-dimensional kinetic terms are,
\eqn\impuritylag{ L_{\rm impurity}  =  \int{dt dx_1 \,\, \left( -
\sum_{\mu=0,1} | D_\mu Q^{f_p}_A(x_1,t) |^2 - \sum_{i=6}^9 | \p_i(x_1, x_p,
t)\, Q^{f_p}_A|^2  \right), }}
where $f_p=1,\ldots,k_p$ and where the connection is evaluated at $x_2=x_p$.
There are two more terms in the potential. The first is proportional to,\foot{
The proportionality constant includes a $\delta (0)$ factor which is essentially
the
order of the symmetry group. This divergent factor is needed to obtain a Higgs
branch with the expected physical properties \rks.}
\eqn\potone{ \sum_{a,A,B}  Q_{Af_p}^{\dagger} T^a Q_{B}^{f_p} 
Q^{Af_p'} T^a Q^{B\dagger}_{f_p'},}
where the $T^a$ are the generators for the fundamental representation. The
final term is the most interesting, and is proportional to,
\eqn\pottwo{\sum_{l,m=1}^4 Q^{Af_p} [ \tp_l, \tp_m] \tau^{lm}_{AB}
Q_{f_p}^{B\dagger} + c.c.}
The fields $ \tp $ are,
\eqn\newfields{ \eqalign{ \tp_1 & = D_2(x_1,x_p,t) \cr
                          \tp_r & = \p_{r+1}(x_1,x_p,t) \quad r=2,3,4.}}
To the localized hypermultiplets, the gauge-field $A_2$ is a scalar filling out
a hypermultiplet that parametrizes the position of the zero-branes in the
longitudinal five-brane. As before, the Coulomb branch of the bulk -- including
the interactions with the localized hypermultiplets -- should describe the
spacetime physics in the background of $k$ type IIB NS five-branes wrapped on
$R_B$. As a function of $R_B$, the matrix model should interpolate between the
tube metric for type IIA NS five-branes and the corresponding metric for type
IIB NS five-branes.

 Let us place all the hypermultiplets at the same point on $x_2$.  The degrees
of freedom relevant for the decoupled physics consist of the hypermultiplets,
$Q_A^f$, the three scalars $\p_3, \p_4, \p_5$ parametrizing motion within the
brane, and a scalar parametrizing motion on the compact direction with radius
$R_B$. The choice of this final scalar depends on the limit we choose for
$R_B$.  Type IIA and IIB NS five-branes wrapped on a longitudinal circle are
equivalent by T-duality. Taking the limit $R_B \r 0$ together with the limits
in \delimits\ should decouple the bulk physics and result in a matrix
description of type IIA NS five-branes. Let us check this is the case. In this
limit, the radius $\S_1$ is fixed but $\S_2$ goes to zero and the theory begins
to look $1+1$-dimensional. As we will check below, in this limit the natural
fourth scalar parametrizing motion on $x_2$ is the scalar $\p_W$ for the Wilson
line around $x_2$. In this limit, $\p_W$ combines with $\p_3, \p_4, \p_5$ into
a $1+1$-dimensional hypermultiplet. The four scalars frozen at the position of
the five-branes combine with the gauge-fields into a $1+1$-dimensional vector
multiplet. To determine whether the effective dynamics is really determined by
a $1+1$-dimensional model, we need to study the dimensionless quantity
\eqn\parameter{ \gamma = g^2_{YM} \S_2 = {1 \over (R_2 M_s)^2},}
in a way analogous to \refs{\rthroat, \rns}. As $R_2 \r \infty$, $\gamma \ll 1$
and theory becomes two-dimensional while the effective gauge interactions are
weak. In this case, it is reasonable to dimensionally reduce to a
$1+1$-dimensional model with $U(N)$ gauge symmetry, an adjoint hypermultiplet,
and $k$ fundamental hypermultiplets. We can check that the period for $\p_W$
really decompactifies in this limit. On reduction, the kinetic term for this
scalar becomes,
$$ {2\pi \S_2 \over g_{YM}^2} \int{ dt dx_1 \, ( {\dot \p_W })^2}, $$
where $ \p_W \sim \p_W + 1/\S_2$. Rescaling the kinetic term to obtain a
dimensionless scalar shows that the period for the circle is proportional to
$1/ \sqrt{\gamma}$ and so decompactifies as $\gamma \r 0$. Since the scalars in
the vector multiplet are frozen, the dynamics corresponding to the decoupled
physics comes from the Higgs branch. The effective $1+1$-dimensional coupling
constant,
\eqn\geff{g_{eff}^2 = g_{YM}^2/ 2\pi \S_2, }
 also goes to infinity. The theory then flows to the conformal field theory of
the Higgs branch in accord with \refs{\rquantumfive, \rstringfive}.

In the opposite limit where $R_2 \r 0$, we obtain type IIB five-branes. Again
$\S_1$ is fixed, while $\S_2 \r 0$. The bulk Yang-Mills coupling is again
driven to infinity.  In this limit where $ \g \gg 1,$ the three-dimensional
gauge dynamics cannot be ignored. Even though $\S_2 \r 0$, we cannot reduce the
model trivially to a $1+1$-dimensional theory. The bulk coupling constant
\ymcoupling\ is driven to infinity so the theory flows to a particular
superconformal field theory for every choice of $N$ and $k>1$.

 In this limit, for the abelian case where $N=1$, the natural scalar to use for
motion on $x_B$ is the dualized gauge-field, $\p_D$, rather than $\p_W$. After
rescaling the kinetic term for $\p_D$, we  see that the period is proportional
to $\sqrt{\g}$. On strictly physical grounds, the direction $x_B$ must become
symmetric with the three non-compact directions of the longitudinal five-brane
as $R_B \r \infty$.  This implies that in the strong coupling limit, the
$Spin(3)$ flavor symmetry is enhanced to $Spin(4)$ with $\p_D$ combining with
$\p_3,\p_4,\p_5$. A similar enhancement must also occur for the non-abelian
case. The use of the dual scalar rather than the scalar for the Wilson line on
$x_2$ played an important role in matrix formulation of the type IIB string.
The chiral spacetime supersymmetries came about because the extra dimension
involved the dual scalar. Here the situation is analogous. In the $R_2 \r
\infty$, we are describing a theory with $(2,0)$ supersymmetry in six
dimensions; in the opposite limit, we obtain a theory with $(1,1)$
supersymmmetry. However, there is an important subtlety in this case. The
localized interactions treat the two gauge-fields $A_1$ and $A_2$ differently.
The interaction breaks Lorentz invariance so the usual dualization procedure
cannot be straightforwardly applied to the hypermultiplet interactions, even in
the abelian case. The strong coupling brane picture in terms of M theory
five-branes and two-branes suggests a possible way of obtaining the dualized
interaction by starting with the membrane action rather than with the D 2-brane
action.

For the case $N=1$, the $2+1$-dimensional gauge dynamics are unlikely to
matter. In this case, all the bulk fields appear quadratically in the
Lagrangian. The coupling can then be scaled into the $1+1$-dimensional quartic
interactions. The decoupled physics should then be governed by a
$1+1$-dimensional sigma model with a target space metric isomorphic to the
metric on the moduli space of one instanton in $SU(k)$ gauge theory. For
example, for $k=2$ the target space would be,
$$ \IR^4 \times \IR^4/\IZ_2.$$
At first sight, this may seem to coincide with the sigma model for the matrix
model for the $(2,0)$ string theory with $N=1$. However, again the $\IR^4$
contains the dual scalar and so the fermion content should differ. It seems
likely that the $\theta$ angle for this orbifold theory is zero since the
theory should not describe free strings in six dimensions \rstringfive. It is
tempting to conjecture that for higher $N$, the decoupled physics will be
described by a sigma model on the moduli space of N instantons in $SU(k)$ gauge
theory. The difference between the $(2,0)$ and $(1,1)$ cases again encoded in
the structure of the fermions. It would be interesting to explore this
possibility further.

\newsec{The Type IIA Matrix String on an ADE Singularity}

To determine the dual realization of the superconformal field theory describing
the decoupled physics, let us study case two. Some related comments have
appeared in \refs{\rfischler, \rHG}. In this description, we have NS
five-branes in type IIB with a compact transverse circle. On T-duality, the
system becomes a configuration of Kaluza-Klein monopoles in type IIA string
theory \refs{\rTduality}. The monopole solution is constructed by taking the
spacetime metric to be a product of a flat metric for five-dimensions with a
multi-Taub-NUT metric \rmultitn. We will briefly recall the main features of
the multi-Taub-NUT metric which have appeared recently in various discussions
of Kaluza-Klein monopoles in string theory and matrix theory
\refs{\rashoke\rgeneralsixbrane -\rmoore, \rHG, \rns}. The metric is determined
by a single function $V$,
\eqn\multitn{ ds^2 = V(x)\, d\vec{x}^2 + V(x)^{-1} (d\theta + \vec{A}
\cdot d\vec{x})^2,}
where $ \grad {} V = \grad {} \times \vec{A}.$ The function $V$ can be written
\eqn\definer{ V = 1  + r \sum_{i=1}^k {1\over |\vec{x}-\vec{x}^i|}. }
 The free parameter $r$ sets the scale for this hyper\kh metric, and the
coordinate $\theta$ has a period proportional to $r$. The $k$ branes are
located at the points $ \vec{x}^i. $ The parameter $r=R_2$ in this case. When
we have $k$ coincident branes, the multi-Taub-Nut has an $A_{k-1}$ singularity
at the location of the coincident branes. Key to this story is the independence
of the decoupled physics from the value of $R_2$ \rns. We are free to take the
limit $ R_2 \r \infty$ which decompactifies the circle coordinatized by
$\theta$ everywhere except at the location of the branes. The space then
becomes $\IR^4/\IZ_k$.

This is fortunate since we do not currently know how to provide a matrix model
description for Kaluza-Klein monopoles in type IIA string theory. Such a matrix
description should be obtained by compactifying a transverse direction, say
$x^9$, to the type IIB NS five-brane. The corresponding matrix model is
$3+1$-dimensional with $2+1$-dimensional hypermultiplets. This theory should be
well-defined since we have compactified only two transverse directions to the
longitudinal five-brane. This is to be contrasted with the matrix model for an
M theory five-brane with three compact transverse dimensions. The matrix model
is similar but the hypermultiplets are not localized. The gauge theory is then
not asymptotically free and needs definition in the ultra-violet. This problem
can also be associated to the deficit angle generated by seven-branes by
T-dualizing the $k$ four-branes on a transverse $T^3$. It is natural to attempt
to define this matrix theory in terms of the six-dimensional theory
corresponding to $N$ $Spin(32)$ instantons compactified on $T^2$, which at low
energies has $Sp(N)$ gauge symmetry. This is the decoupled theory living on
parallel type I five-branes \rsdecoupled. This definition only makes sense for
sixteen or fewer five-branes, and also leads us down the road of `matrix models
for matrix models.' Fortunately, when the hypermultiplets are localized, we
expect no such problems.

For the decoupled physics,  we actually only need a matrix description of the
type IIA string on an $A_{k-1}$ singularity which has been supplied in
\refs{\rdoumoo, \rfischler}\foot{ A slight modification of this original
conjecture has recently been proposed in \rkin, but our results seem to be
compatible with the gauge groups and matter content originally suggested in
\rdoumoo. I wish to thank K. Intriligator for discussions on this point. }. The
matrix theory is a $1+1$-dimensional gauge theory with $(4,4)$ supersymmetry
and gauge group,
\eqn\gaugegroup{ U(N_{i_1}) \times \cdots \times U(N_{i_k}).}
Each factor is associated to a node of the extended Dynkin diagram; see, for
example \refs{\rmirror, \rdoumoo}. For the $A_{k-1}$ case, $N_{i_1}=\ldots =
N_{i_k}=N$. The matter content can also be read from the extended Dynkin
diagram where we obtain a hypermultiplet for each link in the diagram. This can
be summarized in an adjacency matrix whose elements, $a_{ij}$, are one when
there is a link between the $i^{th}$ and $j^{th}$ node and zero otherwise. The
hypermultiplets appear in the representations, $ \oplus_{ij} a_{ij} (N_i, {\bar
N}_j).$ The $k$ hypermultiplets are chosen so that the Higgs branch corresponds
to a product of $A_{k-1}$ singularities. The singularities can be resolved by
turning on Fayet-Iliopoulos parameters. Our interest, however, is not with the
Higgs branch which describes type IIA strings moving on a space-time which is a
product of $ \IR^4$ with an ALE space. When the Fayet-Iliopoulos parameters
are set to zero so the space is singular, a new branch should appear in the
matrix description. Further, the conformal field theory for this branch should
decouple from the spacetime physics. The Coulomb branch of the current matrix
proposal for the type IIA string on an $A_{k-1}$ singularity satisfies these
requirements. It disappears when the singularity is resolved, and it decouples
from the Higgs branch in the infra-red. There are two distinct infra-red limits
because the symmetry flowing to the R charge of the $(4,4)$ superconformal
algebra is necessarily different for the Higgs and Coulomb branches \rwtensor.
On the Coulomb branch, the strings are trapped at the location of the
singularity. That is precisely the picture we desire for the decoupled sector.

The Coulomb branch is a $4Nk$-dimensional space with a metric that has a
tube-like structure. For $N=1$ and $k=2$, the Coulomb branch is one-dimensional
and the metric can be determined essentially by symmetries \rthroat. One $U(1)$
vector multiplet decouples, leaving a $U(1)$ gauge theory with two electrons.
The matrix theory is then a $1+1$-dimensional theory on a circle of radius
$\S_1$ given in \newymdim. In the decoupling limit, the coupling constant for
the gauge theory is driven to infinity. The target space is a product of
$\IR^4$ with a four-dimensional space with metric proportional to $1\over r^2$
and non-trivial torsion \rthroat. We might ask whether it is actually valid to
restrict to the description in terms of a metric on moduli space for small $r$,
since that is roughly a Born-Oppenheimer approximation. A similar argument in
the dimensional reduction of this gauge system to quantum mechanics would seem
to yield a singular $1\over r^3$ metric fixed uniquely by a global $Spin(5)$
symmetry \rglobal. However, there we know that the  description of the physics
in terms of a metric on moduli space breaks down. There is actually a
normalizable ground state wave-function, and the physics is quite non-singular
\rhmon. There is clearly much to be understood about these singular conformal
field theories.

 Now it has been conjectured in \refs{\rwtensor, \rthroat}\ and argued in
\rvafa\ that there should be a dual realization in which the tube is absent.
That dual realization was presented in the previous section, and argued to be a
sigma model on $\IR^4 \times \IR^4/\IZ_2$. Note that there was no torsion in
that realization. The subtlety with the tube metric manifests itself in the
dual description in the guise of a hard to analyze $\theta=0$ orbifold theory.
It is important to stress that the duality that we are claiming follows from a
legitimate string T-duality: the T-duality taking an NS five-brane with a
compact transverse circle to a Kaluza-Klein monopole. This dual realization of
superconformal field theories seems to be the natural analogue of the
three-dimensional mirror symmetry found by Intriligator and Seiberg \rmirror.
It equates a $1+1$-dimensional limit of the `Higgs' branch for a
$2+1$-dimensional theory with localized hypermultiplets with the Coulomb branch
conformal field theory of a $(4,4)$ gauge theory. This duality  extends to the
family of superconformal field theories parametrized by $N$ and $k>1$, and
generally exchanges a metric with torsion for a hyper\kh metric. This duality
is clearly related to conventional two-dimensional mirror symmetry
\rconvmirror, where a theory with $(2,2)$ supersymmetry can often be realized
in terms of sigma models on two or more distinct target spaces. However, in
those cases, the target space is typically compact.

Lastly, we can define the decoupled $(1,1)$ string theory corresponding to the
$D$ and $E$ singularities by using the Coulomb branch of the type IIA matrix
string propagating on those singular spaces. The gauge groups and matter
content have been described in \refs{\rmirror, \rDE}. The hypermultiplets are
chosen so that the Higgs branch is again a product of $D$ or $E$ singularities.
Again the corresponding Coulomb branch conformal field theory will describe the
decoupled physics. For the $D_k$ case, a dual description of the Coulomb branch
conformal field theory naturally follows by considering NS five-branes and
orientifold planes. The $E_6, E_7, E_8$ theories are likely to have more exotic
dual realizations.

\bigbreak\bigskip\bigskip\centerline{{\bf Acknowledgements}}\nobreak

It is my pleasure to thank J. Brodie, K. Intriligator, D. Morrison,  R. Plesser
and particularly O. Ganor, N. Seiberg and E. Witten for helpful discussions.
This work is supported by NSF grant DMS--9627351.

\listrefs
\end